\documentstyle[graphicx,aps,multicol]{revtex}

\begin{document}

\title{From Spectral Relaxation to Quantified Decoherence} 
\author{Chr. Balzer, Th. Hannemann, D. Rei\ss, W. Neuhauser, P.E. Toschek, and
Chr. Wunderlich}

\address{Institut f\"{u}r Laser-Physik, Universit\"{a}t Hamburg, 
Jungiusstr. 9, D-20355 Hamburg, Germany}

\date{August 3, 2001}

\maketitle

\begin{abstract}
Quantum information processing (QIP) requires thorough assessment of
decoherence. Atoms or ions prepared for QIP often become addressed by
radiation within schemes of alternating microwave-optical double resonance.
A well-defined amount of decoherence may be applied to the system when
spurious resonance light is admitted simultaneously with the driving
radiation. This decoherence is quantified in terms of longitudinal and
transversal relaxation. It may serve for calibrating observed decoherence as
well as for testing error-correcting quantum codes.
\end{abstract}

\begin{multicols}{2}
The spectroscopic determination of atomic relaxation is among the earliest
applications of laser spectroscopy [1]. Over the decades that have passed since then,
the corresponding techniques have been dramatically refined, and the demand for
reliable data has expanded. An important aspect is the availability of
respective data derived from observations on individual trapped ions [2].

Laser-cooled ions prepared in an electrodynamic or electromagnetic trap and
individually addressed are convenient building blocks for the storage and
processing of quantum information [3]. In fact, they represent a scalable
system, in contrast with certain competitive approaches, as, e.g., spin
resonance spectroscopy [4]. The more important is the realization and full
understanding of the radiative interaction of such a system: The
implementation of the \textit{coherent} dynamics of an individual atomic
particle being radiatively driven is prerequisite for the demonstration of
any quantum-logical gate, in particular of the basic Hadamard transformation
[5]. The demonstration of the coherently controlled dynamics of trapped ions has
been reported with the vibrational sidebands (or ''one-phonon lines''), and
carrier line (''zero-phonon'') of a laser-driven Raman transition between
ground-state hyperfine levels [6], with a microwave-driven hyperfine carrier
line [7], with the vibrational sidebands of a dipole-forbidden optical line
[8], and with a corresponding optical
carrier line [9]. 

Some of these schemes are troubled by imperfectly
identified kinds of decoherence that would thwart the performance
of a substantial number of successive steps of coherent interaction. This
decoherence may range from the residual decay of the involved metastable
states to spurious light scattering, and to parasitic phase fluctuations of
the applied radiation. For instance, the decoherence on the vibrational
dynamics of ions localized in a Paul trap [5] has been suggested recently to
result from parasitic spontaneous scattering of the two off-resonant light
fields that serve as the pump and Stokes waves for Raman excitation of the
hyperfine resonance [10]. Also, various types of reservoirs have been
implemented that make the specific motional states decay in characteristic
ways [11].

In a microwave-optical double-resonance experiment, we have found  
decoherence to appear during the coherent evolution of an individual trapped and
cooled $^{171}Yb^{+}$ion that was microwave-driven on the ground-state hyperfine
transition. The evolution was monitored via the corresponding spin nutation.
For this purpose, the ion was \textit{alternatingly} driven and probed by
highly monochromatic microwave radiation, and by resonant scattering of
laser light, respectively. The observed decoherence was brought about by a
minute amount of residual laser light that impinged upon the ion while the
microwave drive was applied. Blocking out this spurious light eliminated the
decoherence. 

\begin{figure}
\begin{center}
\includegraphics[width=7cm]{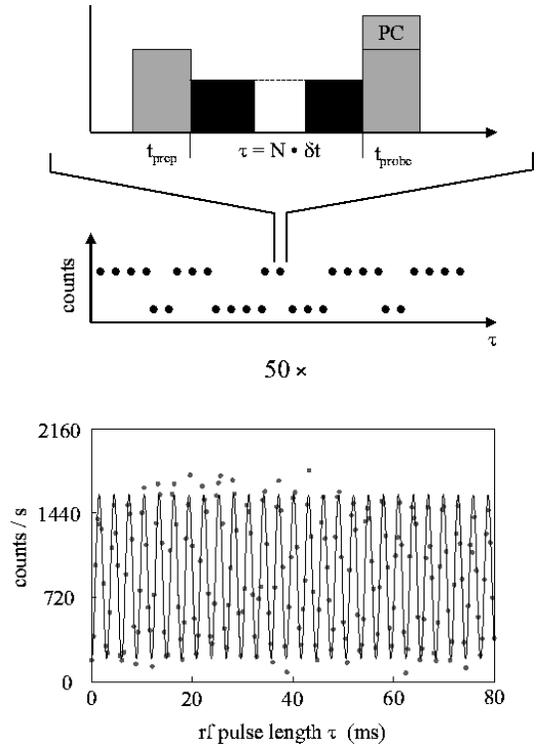}
\caption{An individual measurement includes a pulse of resonant light
that prepares the ion in the $F=0$ state of the ground-state hyperfine
doublet, a microwave driving pulse that is $N$ times a unit length $\delta t$%
, and a probe light pulse. The scattered light is detected by a photon
counting (PC) system (top). With each subsequent measurement, $N$ is
incremented by a unit, up to $N_{\max }$, and yields a sequence of $N_{\max
} $ data, ''on'' or ''off'' (center). Accumulation of 50 sequences reveals
rf nutation on the ground-state hyperfine transition (bottom).}
\end{center}
\end{figure}

We shall show that well-defined quantities of
longitudinal and/or transversal relaxation may be admitted to the ion by
controlling the strength and polarization of the light and the level and
direction of the ambient magnetic field. Such a '`designed" decoherence
seems useful for quantitative comparison with observed decoherence and as a
testing ground for the sensibility of quantum algorithms to the action of
decoherence, as well as for the function of error-correcting codes.

The double-resonance experiment was performed on a single $^{171}Yb^{+}$ion%
\textbf{\ }in a 2 mm-sized electrodynamic trap. The $F=0\rightarrow 1$
hyperfine resonance of the ion's $S_{1/2}$ ground state was driven, for $%
\delta t=$ 100 or 400$\mu $s, by 12.6 GHz microwave radiation. Probing the $%
F=1$ state immediately followed, when the ion was illuminated by a 5 ms pulse of
369 nm laser light. The excitation and observation of resonance scattering
on the $S_{1/2}(F=1)\rightarrow P_{1/2}(F=0)$ line proves the
preceding attempt of microwave excitation to have succeeded; the absence of scattering proves
this attempt to have failed. A frequency-doubled Ti:sapphire laser generated
369 nm light of about 100 kHz bandwidth. This light was scattered on the
ion's $S_{1/2}-P_{1/2}$ resonance line, such that the ion was cooled deep
into the Lamb-Dicke regime. Occasional optical pumping of the ion into its
metastable $D_{3/2}$ level was counteracted, by using 935 nm light from a diode
laser to repump the ion into the ground state.

Trajectories of measurements
have been recorded, each of which includes an initial preparatory light pulse
that pumps the ion into state $F=0$, a microwave driving pulse of length $%
\tau =$ N$\delta $t, a probe-light pulse, and simultaneous recording of the
presumptive fluorescence scattered off the ion (Fig. 1). Within a
trajectory, N varied from 1, indicating the initial measurement, through
300, the final one. The recorded signals form quasi-random sequences of
'`on'' and '`off'' results; however, the superposition of many accumulated
trajectories yields the probability $P_{1}(\theta )$ for occupation of the $%
F=1$ ground level by the ion, as a function of the driving time $\tau $. The
data of 50 trajectories have been superimposed, and modulation of $P_{1}$
emerges that is ascribed to Rabi nutation of the ion driven by the microwave
pulses of area $\theta =\Omega \tau $, whose duration $\tau $ was stepwise
extended (Fig. 1, bottom).

\end{multicols}{2}

\begin{figure}
\begin{center}
\includegraphics[width=12cm]{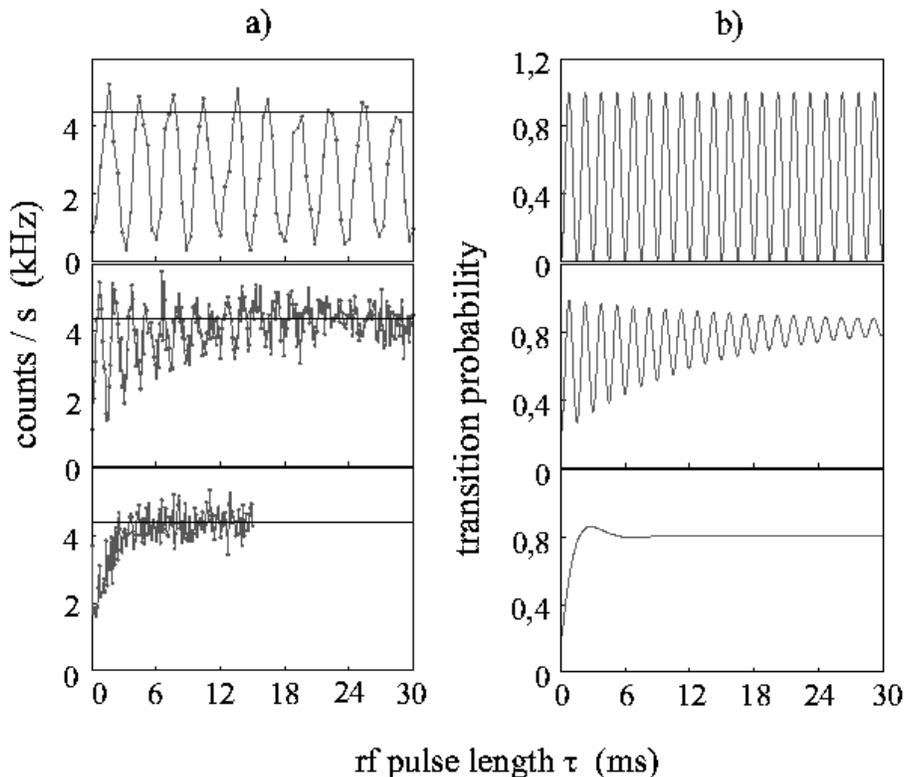}
\caption{Scattered-light response vs. length of microwave driving
pulse. (a) Observed, with level of light applied to the ion \textit{during}
the drive: no light (top), 2nW (center), 20nW (bottom). (b) Simulated: Rabi
frequency $\Omega _{l}=0$ (top), 50 kHz (center), 500 kHz (bottom).}
\end{center}
\end{figure}

\begin{multicols}{2}

The nutational oscillation shows high contrast. However, when a spurious
level of probe light is admitted to the ion \textit{simultaneously} with the
driving pulse, the evolution of the probability of light scattering is
dramatically modified (Fig. 2a): (i) the Rabi nutation becomes damped with
its time constant being reduced upon increased intensity of the background
light, and (ii) at long times $\tau ,$ the probability $P_{1}(\theta )$
saturates to a level somewhere between 1/2 and unity, and not
necessarily to level 1/2.

In order to account for these observations, the ion, the light fields, and
their interaction have been modelled by four-level Bloch equations (Fig.
2b). The corresponding level scheme is shown in Fig. 3. The Zeeman sublevels 
$m_{F}=+1$ and $-1$ have been combined and make up for the effective level 2
not affected by the driving microwave radiation on the hyperfine transition
0-1, with $\Delta m_{F}=0$.

\begin{figure}
\begin{center}
\includegraphics[width=6.5cm]{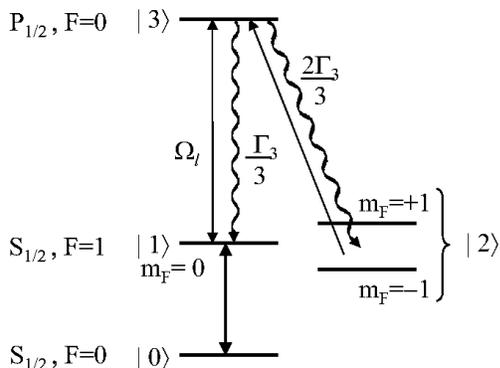}
\caption{Simplified level scheme of the $^{171}$Yb$^{+}$ ion used in
the microwave-optical double resonance experiment.}
\end{center}
\end{figure}

The minute light intensity applied to the ion during the driving intervals
is far below saturation of the resonance line, $I=\Omega _{l}^{2}/\gamma
_{l}\Gamma _{3}\ll 1,$ where $\Omega _{l}$ is the Rabi frequency of the ion
generated by the laser light, $\Gamma _{3}$ and $\gamma _{l}=\Gamma
_{3}/2+\gamma _{lph}$ are the constants of energy relaxation of the
resonance level $(P_{1/2}),$ and of the phase relaxation of the laser-excited
(electric) dipole, respectively. In the present experiment, the extra rate
is small, $\gamma _{lph}\ll \Gamma _{3},$ and it is neglected. In the above
limit it is appropriate, for the interpretation of the ionic evolution, to
restrict the modelling of the \textit{optical} part of the dynamics --- ,
i.e. the part related to the three levels 1, 2, 3 --- to optical pumping and
decay, in terms of rate equations. This rate evolution is coupled to the
microwave-driven coherent evolution, on the two-level system 0-1, via level
1 that represents state $S_{1/2}(F=1,$ $m_{F}=0).$ Now, the latter coherent
dynamics is affected by light-generated decoherence of two kinds: (i) a net
loss of population, from the driven two-level system 0-1, by optically
pumping the ion into state 2, i.e. the $F=1$ Zeeman sublevels $m_{F}=\pm $ $%
1$, with subsequent repumping, and (ii) additional loss of phase coherence
of the driven spin dynamics by Rayleigh scattering into the eigenstate $1$,
i.e. $F=1,$ $m_{F}=0.$

The loss of some population from state 1 to state 2 makes the probability $%
P_{1}$ of finding the ion in the entire probed hyperfine state $F=1$ (states
1 and 2) saturate \textit{above} 1/2, since the Zeeman sublevels $\pm 1$
retain part of the population from the coherent transfer --- on the
condition $\Delta m_{F}=0$ --- into state $F=0$. This fractional population
may become re-excited, however, by a component of the linearly polarized
laser light and taken back to state 1 ($F=1,$ $m_{F}=0)$ during the interval
of probing, since the width of the resonance line far exceeds the small
Zeeman splitting of the $S_{1/2}(F=1)$ state. Complete pumping to level 2 $%
(m_{F}=\pm 1)$ makes $P_{1}$ saturate at unity. The scattering rates are $%
\beta _{f\text{ }}n_{i}r_{i}$, where $n_{i}$ is the population in the
initial state, $\beta _{f}$ is the branching ratio of the decay of state 3
into the final state $(\beta _{1}=\frac{1}{3},$ $\beta _{2}=\frac{2}{3}),$
and the scattering rate per atom is the average population $\langle
P_{3}\rangle $ of state 3, times the decay rate $\Gamma _{3},$ such that
\begin{equation}
r_{1}=\langle P_{3}(0)\rangle \Gamma _{3}  \label{1}
\end{equation}
\begin{equation}
r_{2}=\left( \langle P_{3}(+1)\rangle \text{ }+\text{ }\langle
P_{3}(-1)\rangle \right) \text{ }\Gamma _{3}\text{ .}  \label{2}
\end{equation}
Here ($m\equiv m_{F})$, 
\begin{equation}
\langle P_{3}(m)\rangle \text{ }=\frac{1}{2}\text{ }\frac{I(m)L(B,m)}{%
1+I(m)L(B,m)}\text{ ,}  \label{3}
\end{equation}
where $I(\pm 1)=I_{0}\sin ^{2}\alpha $, $%
I(0)=I_{0}\cos ^{2}\alpha ,$ $I_{0}$ is the density of light flux at the
ion's location, $\alpha $ is the angle subtended by the direction of the
light polarization and
magnetic field $B$. $L(B,m)$ is defined by
\[
L(B,m)=\frac{(\Gamma _{3}/2)^{2}}{(\Gamma _{3}/2)^{2}+(\omega
_{0}-\omega +m\delta )^{2}}\text{ ,} 
\]
$\omega $ and $\omega _{0}$ are the light and resonance
frequencies, respectively, $\delta =g_{F}\mu _{B}B/\hbar $, $g_{F}$ $\simeq
1 $ is the Land\'{e} factor of state 2, and $\mu _{B}$ is the Bohr magneton [12].
Now, let us further restrict the model to the two-level system made up by
states 0 and 1, and attribute to it conventional rates per atom of phase and
energy relaxation, $\gamma $ and $\Gamma $, respectively. These rates will
become \textit{identified }with quantities taken from the above three-level
model: the spin nutation gives rise to oscillation of the population
difference in the two-level system that is supposed to damp out
exponentially with an effective constant $\gamma $ of phase relaxation [12],
such that

\begin{equation}
\gamma =\frac{\Gamma }{2}+\gamma _{ph}\text{ }\hat{=}\text{ }r_{1}\text{ ,} 
\label{4}
\end{equation}
where $\gamma _{ph}$ is the contribution of extra phase
perturbation not related to intrinsic relaxation. Moreover, the flow
equilibrium established by the scattering as well as the quasi-steady state
on the microwave-driven line after dephasing require
\begin{equation}
n_{0}=n_{1\text{ }}=\frac{1-n_{2}}{2}  \label{5}
\end{equation}
and
\begin{equation}
\frac{1}{3}n_{2\text{ }}r_{2\text{ }}=\frac{2}{3}n_{1\text{ }}r_{1\text{ }}.
\label{6}
\end{equation}
Thus, $n_{2}=\frac{r_{1}}{r_{1}+r_{2}}$ , and the normalized probe
signal, i.e., the probability of finding the system in the upper state 1 of
a relaxing two-level system is
\begin{eqnarray}
P_{1}^{(3)} &=&n_{1}+n_{2}\text{ }=\text{ }\frac{1+n_{2}}{2}  \label{7} \\
&=&1-\frac{1}{2}\text{ }\frac{r_{2}/r_{1}}{1+r_{2}/r_{1}}  \nonumber \\
&=&1-P_{0}^{(3)}\text{ .}  \nonumber
\end{eqnarray}

 One may identify $P_{0}^{(3)}$ with an effective two-level
excitation probability $P_{1}^{(2)}=\frac{1}{2}$ $\frac{\text{I}}{1+\text{I}}
$, where I$\ =\Omega ^{2}/\Gamma \gamma $, and $\Omega $ is the Rabi
frequency of the microwave. This interpretation (i) yields the effective
rate constant of energy relaxation,
\begin{equation}
\Gamma \text{ }\hat{=}\text{ }\frac{\Omega }{r_{2}}\text{ ,}  \label{8}
\end{equation}
and (ii) shows that this effective relaxation makes the ion decay
into the \textit{excited} state 1. Note that the two effective rate
constants $\gamma $ and $\Gamma $ of eqs. 4 and 8 may be set \textit{%
separately} by suitable selection of the light polarization and/or of the
ambient magnetic field. Fig. 4 shows trajectories calculated with selected
values of $T_{0}$ $=\Gamma ^{-1}$ and $T_{2}=\gamma ^{-1}.$ They demonstrate
various degrees of damping as well as different levels of saturation of $%
P_{1}(\tau ).$

With a single atom, competing spontaneous decay into non-degenerate levels
may preserve coherence in the atom [13]. Therefore, complete modelling of
the rate of optical pumping represented by eq. 2 would require one to
include an interference term of the transition amplitudes of the
back-and-forth optical pumping, via the states $F=1,$ $m_{F}=+1$ and $-1$,
that give rise to indiscernible pathways. This term displays a resonance in
zero magnetic field and represents what is called ''zero-field level
crossing'' [14], a ground-state Hanle effect. The width of the crossing
resonance is determined by the lifetime of the intermediate interfering
states. The $F=1$ levels show an effective lifetime on the order of
milliseconds, or longer. Thus, the interference term in rate $r_{2}$ would
make this rate vary across a spectral tuning range of the laser less than
1 kHz wide. This spectral feature is not resolved by the emission bandwidth
of the laser.
\begin{figure}
\begin{center}
\includegraphics[width=8cm]{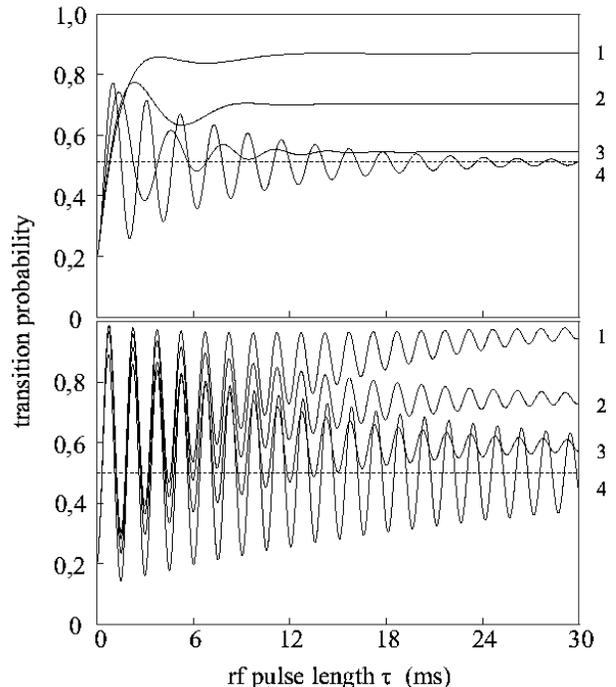}
\caption{ Probability $P_{1}(\theta )$ of the ion to be found in the
upper state 1 after having been driven by a microwave pulse of area $\theta
=\Omega \tau .$ Weak resonance scattering excited by laser light of
intensity $\ I=\Omega _{l}^{2}/\Gamma _{3}\gamma _{l}$ mimics phase
relaxation and energy decay into state 1. Initial population in 0: 0.8, in
1: 0.2. Detuning $\delta =-0,28,$ $\delta _{l}=-3\times 10^{3}.$ $\Gamma
_{3}=18\times 10^{3},$ $\Omega =4,2.$ Top: $\sqrt{2r_{1}\gamma _{l}}=700,$ $%
\sqrt{r_{2}\gamma _{l}}=70(1),$ $140(2),$ $350(3),$ $700(4).$ Bottom: $\sqrt{%
2r_{1}\gamma _{l}}=70,$ $\sqrt{r_{2}\gamma _{l}}=7(1),$ $70(2),$ $140(3),$ $%
350(4).$ All frequencies in $2\pi $ $\times $ kHz.}
\end{center}
\end{figure}

Historically it is interesting, that the Hanle effect when observed on Ne
atoms irradiated by the light field of a HeNe laser was one of the earliest
laser-spectroscopic topics; it was dealt with in the diploma thesis of
Theo H\"{a}nsch [15,16].

The availability of easily quantifiable longitudinal and transversal
relaxation that is \textit{light-induced} upon individual atomic systems
displays important advantages when it comes to the application of such a
system to manipulations in QIP impaired by loss of coherence. In particular,
codes of information processing may be tested for their applicability under
this challenging but commonplace condition. On the other hand, codes for
error correction may be made to demonstrate their capacity upon increasing
levels of decoherence fed into the system. The light-induced decoherence, as
demonstrated in this report, is readily applicable to individually addressed
quantum systems, it may be switched on and off immediately, and it is
reproducible.

In summary, a simple method has been outlined that adds a predetermined
degree of decoherence on the coherent radiative interaction of an individual
atom or ion that is to be used for information processing. Although, in the
present experiment, the coherent drive was microwave radiation resonant with
a ground-state hyperfine transition, the same principle seems to apply to a
system where a dipole-forbidden optical transition is driven by laser light.

This work was supported by the Deutsche Forschungsgemeinschaft and the
Hamburgische wissenschaftliche Stiftung.

\end{multicols}{2}
\end{document}